\RequirePackage[figuresright]{rotating}
\documentclass[format=acmsmall,review=false,screen=true]{acmart}
\usepackage{graphicx}
\usepackage{amsmath}
\usepackage{mathtools}
\usepackage{comment}
\usepackage[subrefformat=parens,labelformat=parens]{subfig}
\usepackage{bm}
\usepackage{multirow}
\usepackage{threeparttable,booktabs}
\usepackage{blkarray}
\usepackage{tikz}
\usepackage{balance}
\usepackage{courier}                                       
\usepackage{cleveref}                                      
\usepackage[mathcal]{eucal}
\usepackage[]{algpseudocode}                               
\algrenewcommand\textproc{\texttt}
\makeatletter\let\float@addtolists\relax\makeatother
\usepackage{algorithm}
\usepackage{filecontents}                                  
\usepackage{pgfplots}
\usepackage{pgfplotstable}
\pgfplotsset{compat=newest}
\usepackage[figuresright]{rotating}

\theoremstyle{plain}

\theoremstyle{definition}
\newtheorem{mydefinition}{\textbf{Definition}}

\usepackage[skip=1pt]{caption}            
\graphicspath{{./figs/}}

\begin{document}

\title{
Task modules Partitioning, Scheduling and Floorplanning for Partially Dynamically Reconfigurable Systems  Based on  Modern Heterogeneous FPGAs 
}

\author{Bo Ding}
\affiliation{
    \department{Department of Electronic Science and Technology}
    \institution{University of Science and Technology of China}
    \city{Hefei}
    \country{China}
}
\email{dingbo@mail.ustc.edu.cn}

\author{Jinglei Huang}
\affiliation{
	\department{Department of Electronic Science and Technology}
	\institution{University of Science and Technology of China}
	\city{Hefei}
	\country{China}
}
\email{huangjl@mail.ustc.edu.cn}

\author{Junpeng Wang}
\affiliation{
	\department{Department of Electronic Science and Technology}
	\institution{University of Science and Technology of China}
	\city{Hefei}
	\country{China}
}
\email{wjp97@mail.ustc.edu.cn}

\author{Qi Xu}
\affiliation{
	\department{Department of Electronic Science and Technology}
	\institution{University of Science and Technology of China}
	\city{Hefei}
	\country{China}
}
\email{xuqi@ustc.edu.cn}

\author{Song Chen}
\affiliation{
	\department{Department of Electronic Science and Technology}
	\institution{University of Science and Technology of China}
	\institution{Institute of Artifcial Intelligence, Hefei Comprehensive National Science Center}
	\city{Hefei}
	\country{China}
}
\email{songch@ustc.edu.cn}

\author{Yi Kang}
\affiliation{
	\department{Department of Electronic Science and Technology}
	\institution{University of Science and Technology of China}
	\institution{Institute of Artifcial Intelligence, Hefei Comprehensive National Science Center}
	\city{Hefei}
	\country{China}
}
\email{ykang@ustc.edu.cn}

\begin{abstract}
	Modern field programmable gate array(FPGA) can be partially dynamically reconfigurable with heterogeneous resources distributed on the chip. And FPGA-based partially dynamically reconfigurable system(FPGA-PDRS) can be used to accelerate computing and improve computing flexibility. 
	However, the traditional design of FPGA-PDRS is based on manual design.    
	Implementing the automation of FPGA-PDRS needs to solve the problems of task modules partitioning, scheduling, and floorplanning on heterogeneous resources.
	Existing works only partly solve problems  for the automation process of FPGA-PDRS or model homogeneous resource for FPGA-PDRS. 
	To better solve the problems in the automation process of FPGA-PDRS and narrow the gap between algorithm and application, in this paper, we propose a complete workflow including three parts, pre-processing to generate the list of task  modules  candidate shapes according to the resources requirements, exploration process to search the solution of task modules partitioning, scheduling, and floorplanning, and post-optimization to improve the success rate of floorplan.	
	Experimental results show that, compared with state-of-the-art work, the  proposed complete workflow can improve performance by 18.7\%, reduce communication cost by 8.6\%, on average, with improving the resources reuse rate of the heterogeneous resources on the chip. And based on the solution generated by the exploration process,  the post-optimization can improve the success rate of the floorplan by 14\%.	
\end{abstract}



\maketitle


\section{Introduction}

	With the emergence of intelligent fields such as big data, cloud computing, edge computing,  neural network,  and others,  the applications may be compute-intensive or memory-intensive and therefore put forward increasingly requirements on the ability of hardware platforms \cite{liu2019survey}. 
	During so many hardware platforms, the platforms that can be reconfigurable  have drawn much attention from industry and academia because of possessing lots of superiority in the design of accelerators\cite{wang2017reconfigurable}. 	
	The reconfigurable hardware platform FPGA  has the  characteristics of software definition and hardware realization, has high flexibility, and can significantly shorten the design cycle compared to ASIC \cite{vipin2018fpga}.
	Therefore, FPGA with reconfigurable technology can be the hardware platform to face the challenge that various computing applications may be compute-intensive or memory-intensive.
	
	Traditionally, we design and implement circuits and then deploy these circuits on FPGA, which is a static way to  speed up applications. When the FPGA is running, we have to power off the FPGA and configure it to new circuits if we need to perform new functions. 
	With the development of FPGA technology, dynamic reconfiguration (DR) has been developed with increasing flexibility. 
	DR brings higher-level automation to reconfigure FPGA by changing predetermined functions at run-time. 
	However,  the entire chip must be reconfigured to implement  new circuits each time with DR technology.  
	Thus, significant reconfiguration overhead is incurred because of loading the configuration file each time\cite{hauck2010reconfigurable}.
	Several techniques have emerged to reduce the impact of reconfiguration overhead, such as module reuse, configuration prefetching, and partial dynamic reconfiguration(PDR).  
	During these technologies, PDR can optimize performance, reduce power consumption, and improve design flexibility because  PDR technology can partition the chip into several reconfigurable regions and configure them into corresponding circuits. These regions can be executed in parallel without affecting each other\cite{koch2012partial}. 
\subsection{Related Works}
\label{sec:related_works}	
	
	The design automation of partially dynamically reconfigurable systems involves the problems of  task modules partitioning, scheduling, and floorplanning on heterogeneous resources.
	The analysis of these problems is described in Section \ref{sec:problem_analy}.
	Many previous works have been committed to designing algorithms to implement the automation of partially dynamically reconfigurable systems.  
	However, these works are more or less impaired. 
	And Table \ref{table:related-works} lists part representative works.
	
	
	Some works have focused on partition and floorplan of PDRS to decrease wire length and improve resources utilization.  
	Bolchini et al. \cite{bolchini2011automated} proposed a floorplanner  to place reconfigure regions(RRs). This work is based on the SA algorithm,  and the objective is to improve the utilization  of heterogeneous resources  of the PDRS with optimizing  wire length. However, this work only considers the floorplan of reconfigurable regions and only solves part problems of the design automation of PDRS.
	P. Banerjee et al.\cite{banerjee2010floorplanning} proposed a three-stages heuristic method to partition and floorplan task modules on heterogeneous resources of PDRS. This work limits that each reconfigurable region deploys one task module in each partition.   However, on a modern FPGA chip with PDR  technology, each reconfigurable region can deploy multiple task modules, thus reducing control and communication costs and avoiding congestion. 
	Yao et al. \cite{yao2022fast} proposed the fast search and efficient  algorithm for the floorplan of task modules on modern heterogeneous FPGAs with increasing both floorplan speed and  success rate. However, this work only considers the cost of resource utilization and ignores other costs, such as the cost of wire length.
	These works \cite{bolchini2011automated,banerjee2010floorplanning,yao2022fast} do not consider the task modules scheduling that affects the performance of PDRS.   Besides, some other excellent works also only consider the partition and floorplan in PDRS but don't consider schedule. \cite{rabozzi2015floorplanning,liu2013resource,montone2010placement,singhal2006multi}.
	
	And some other works have focused on partition and schedule. 
	Ganesan et al. \cite{ganesan2000integrated}  proposed a reconfigurable processor system with two reconfigurable regions for performance acceleration.  This work exploits the characteristics of PDR to hide the configuration overhead and improve the parallelism to speed up. However, modern FPGA chips with PDR technology can support multiple reconfigurable regions, and thus this problem is more complicated.
	Rana et al.\cite{rana2009minimization} proposed a reconfigurable overhead minimization method based on the analysis of the communication graph. This work tries to group task modules that require simultaneous reconfguration into the same reconfigurable region. However, the designer must appoint the number of the reconfigurable regions manually. R. Cordone et al. \cite{cordone2009partitioning} proposed an integer linear programming (ILP) model and a heuristic method for task graph partitioning and scheduling. They exploit the characteristics of task module reuse and configuration prefetching to minimize reconfiguration overhead. \cite{jiang2007temporal,purgato2016resource,tang2020partitioning} also contribute to task module partitioning and scheduling in PDRS.
	
	However, task modules partitioning affects schedule and floorplan, and these three factors significantly impact the performance, resources utilization, and energy consumption of PDRS. And some works have focused on solving the problems of task modules partitioning, scheduling, and floorplanning of PDRS integrated. 
	P. Yuh et al. \cite{yuh2004temporal,yuh2007temporal} modeled the task modules as three-dimensional (3D) boxes and proposed an SA-based 3D floorplanner to floorplan and schedule task modules. However, assuming task modules  can be reconfigured at any time and in any region may not match practically reconfigurable architectures.
	E. A. Deiana et al.\cite{deiana2015multiobjective} proposed a scheduler based on mixed-integer linear programming (MILP). The scheduler can schedule and deploy applications on PDRS based on FPGA. And if the scheduler cannot successfully perform, the scheduler will be re-executed until a feasible ﬂoorplan is determined.
	However, the time consumption of the MILP-based approach is impractical for large-scale applications. 
	Chen et al. \cite{chen2018integrated} proposed the partitoned sequence triple \textit{P-ST} to express  partition, schedule, and floorplan of task modules for PDRS and used an SA-based search engine with a perturbation method to achieve a higher success rate of floorplan with optimizing schedule time. However, this work models the process of PDRS by modeling homogeneous resource, CLB, and thus simplifies these problems.
	
	These works mentioned above either solve the three problems sequentially or solve only two of the three problems in an integrated framework, or model the process of PDRS by modeling homogeneous resources, thus simplifying these problems. 
\begin{table}[htbp]
	\centering\small
	\caption{Part representative works related to PDRS}  
	\label{table:related-works} 
	\begin{threeparttable}
	\begin{tabular}{lllllll}  
		\hline 
		Works													&Object	&Partition	&Floorplan	&Schedule	&Method		&Model	\\
		\cite{bolchini2011automated}							&RR		&			&$\surd$	&			&SA-based	&Hetero  \\
		\cite{banerjee2010floorplanning}\cite{yao2022fast} 		&Task	&$\surd$	&$\surd$	&			&Heuristic	&Hetero  \\		
		\hline 
		\cite{ganesan2000integrated}\cite{rana2009minimization}	&Task	&$\surd$	&			&$\surd$	&Heuristic	&Hetero  \\			
		\cite{cordone2009partitioning}							&Task	&$\surd$	&			&$\surd$	&Heur/ILP	&Hetero  \\	
		\hline 
		\cite{yuh2004temporal}\cite{yuh2007temporal}					&Task	&			&$\surd$	&$\surd$	&SA-based	&Homo  \\	
		\cite{deiana2015multiobjective}							&Task	&			&$\surd$	&$\surd$	&MILP	&Hetero  \\
		\cite{chen2018integrated}								&Task	&$\surd$	&$\surd$	&$\surd$	&SA-based	&Homo	\\
		\hline
		\textbf{Our work}										&Task	&$\surd$	&$\surd$	&$\surd$	&Enum/SA/ILP	&Hetero \\
		\hline	
	\end{tabular}
	\begin{tablenotes}
		\item RR/Task:Reconfigurable region/Task module.
		\item Homo: Model only one  kind resource, CLB.
		\item Hetero: Model several kinds  resources among CLB, BRAM, DSP.
	\end{tablenotes}
	\end{threeparttable}	
\end{table}


\subsection{Our Contributions}
\label{sec:contributions}
	In this paper,  we propose a complete workflow to find the optimal solution of task modules partitioning, scheduling, and floorplanning on heterogeneous resources based on FPGA-PDRS.  This workflow includes pre-processing,  the exploration of task modules partitioning, scheduling, floorplanning, and post-optimization.  
	The main contributions of this paper are outlined as follows.
	
\begin{enumerate}
	\item[1)] In this pre-processing process, we propose the strategy of generating the list of task  modules  candidate shapes according to resources requirements of each task module. This strategy is based on the enumeration of task module shapes.
	\item[2)] We change our previous algorithm\cite{chen2018integrated} by  changing the process of exploring solution space and adding the cost of heterogeneous resource utilization to generate the solution of task modules partitioning, scheduling and floorplanning.  
	Then we  integrate the changed algorithm into the complete workflow. 
	\item[3)] After given the partition, schedule, and floorplan of task modules, 
	we build the ILP model to reselect task modules shape from the  candidate shapes list to improve the success rate of floorplan. 

\end{enumerate}

The experimental results demonstrate the efficiency and effectiveness of the proposed algorithms.
The remainder of the paper is organized as follows. 
Section~\ref{sec:problem} not only describes the  hardware architecture and the process of FPGA-PDRS  but also describes the definition   and  analysis of problems.
Section~\ref{sec:representation-workflow} discusses the overall design workflow of  generating  the partition, schedule, and floorplan of task modules  for FPGA-PDRS.
Section~\ref{sec:Algorithms} describes the proposed algorithms  of pre-processing  and  post-optimization.
The experimental results and conclusions are shown in Section~\ref{sec:experiments} and Section~\ref{sec:conclusion}, respectively.

\section{Problem Description}
\label{sec:problem}

\subsection{FPGA-based Partially Dynamically Reconfigurable Architecture}
\label{sec:architecture}

\subsubsection{\textbf{Heterogeneous resources on FPGA chip}}
\label{sec:resouces_structure}

	In modern FPGAs, the mainstream products are Xilinx devices. And there are mainly three kinds of programmable  resources on an FPGA chip, configurable logic block (CLB), block random access memory (BRAM), and digital signal processing (DSP) unit. 
	
\begin{figure}[htbp]
	\centering
	\includegraphics[width=0.48\textwidth]{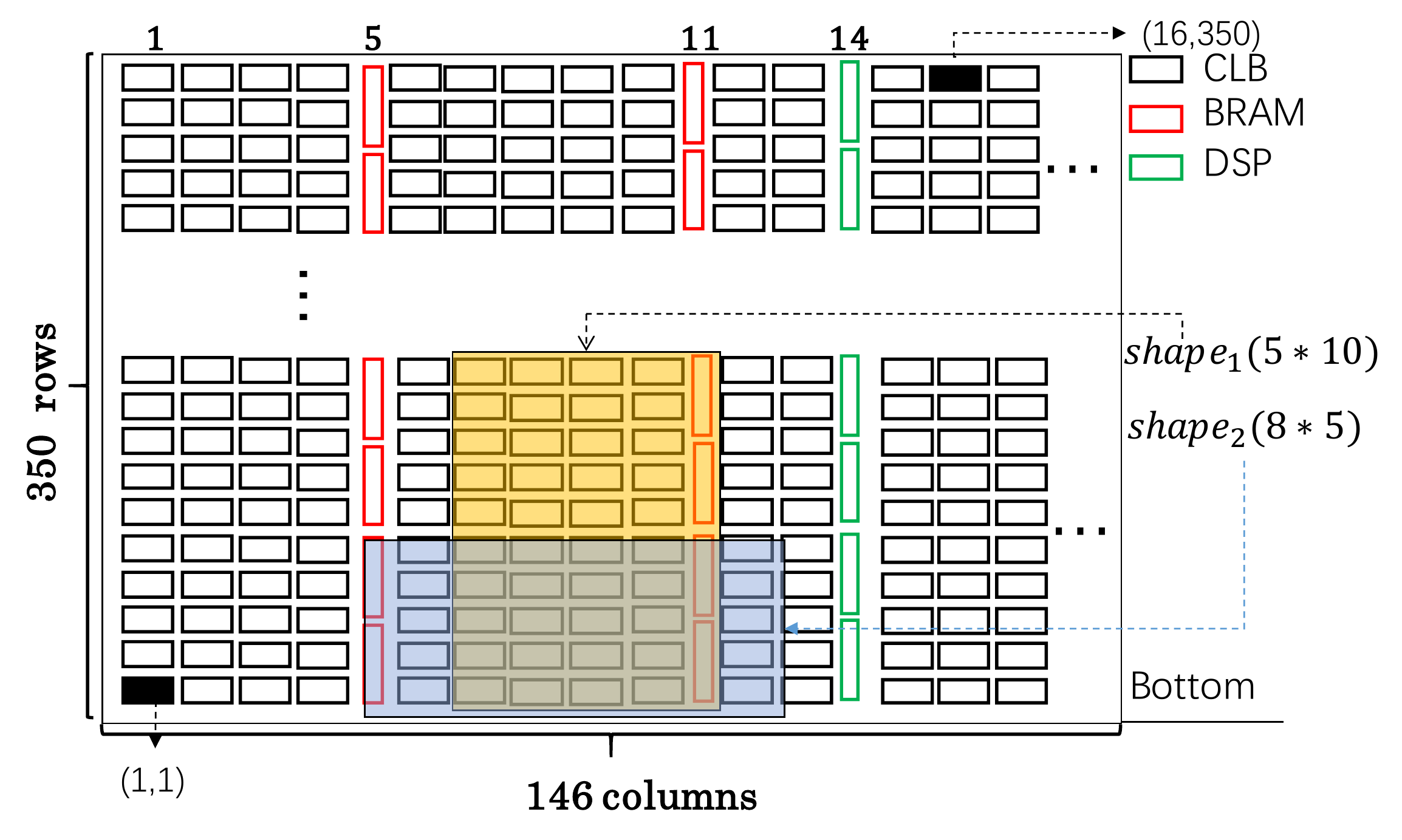}
	\caption{The column structure of heterogeneous resources on the FPGA chip (Eg.XC7VX485T).}
	\label{fig:FPGA_resources}
\end{figure}

	As shown in Fig. \ref{fig:FPGA_resources}, these heterogeneous resources are organized into irregular columns and  distributed on the chip.  Fig. \ref{fig:FPGA_resources} use the chip $XC7VX485T$  as an example \cite{vc707}. $XC7VX485T$  is the chip of evaluation kit VC707, which belongs to the family of Xilinx Virtex-7 series. In $XC7VX485T$, there exists 146 column heterogeneous resources, including 111 columns of CLB, 15 columns of BRAM and 20 columns of DSP.  Each column resource of CLB has 350 rows of tiles, while each column resource of BRAM or DSP has only 140 rows of tiles.

\begin{equation}
	\label{COl_{BRAM}}
	\begin{aligned}
		&Col_{BRAM} = \{5, 11, 23, 29,37,48,66,77,88,99,110,118,124,136,142\}	
	\end{aligned}
\end{equation}

\begin{equation}
	\label{COl_{DSP}}
	\begin{aligned}
		\begin{split}
			&Col_{DSP} = \{14,20,26,34,40,45,51,63,69,74,80,85,91,96,102,107,113,121,127,133\}	
		\end{split}
	\end{aligned}
\end{equation}

To model the heterogeneous resources of an FPGA chip, we use $Col_{BRAM}$ and $Col_{DSP}$ to record the column position of the resource BRAM and DSP, respectively.  For $XC7VX485T$, the value of $Col_{BRAM}$ and $Col_{DSP}$ is shown by Equation (\ref{COl_{BRAM}}) and Equation (\ref{COl_{DSP}}), respectively.
Then we can define the coordinate of each tile.  For example, the coordinate of tile filled with black is $(1,1)$ and $(16,350)$, respectively in  Fig. \ref{fig:FPGA_resources}.

\subsubsection{\textbf{The process of partially dynamically reconfiguration based on FPGA}}
\label{sec:process_PDR}
  	On FPGAs with PDR technology, the  device can be partitioned into several reconfigurable regions. 
  	Each reconfigurable region can be reconfigured into different circuits at different time intervals.  Here regard these time intervals of a reconfigurable region as time layers.
  	To partition each task module into the related reconfigurable region and determine the related configurable time is highly complex. However, there is no complete automation tool to settle this problem, and the design of a partially dynamically reconfigurable system still depends on the human experience.
  	
\begin{figure}[htbp]
	\centering
	\includegraphics[width=0.48\textwidth]{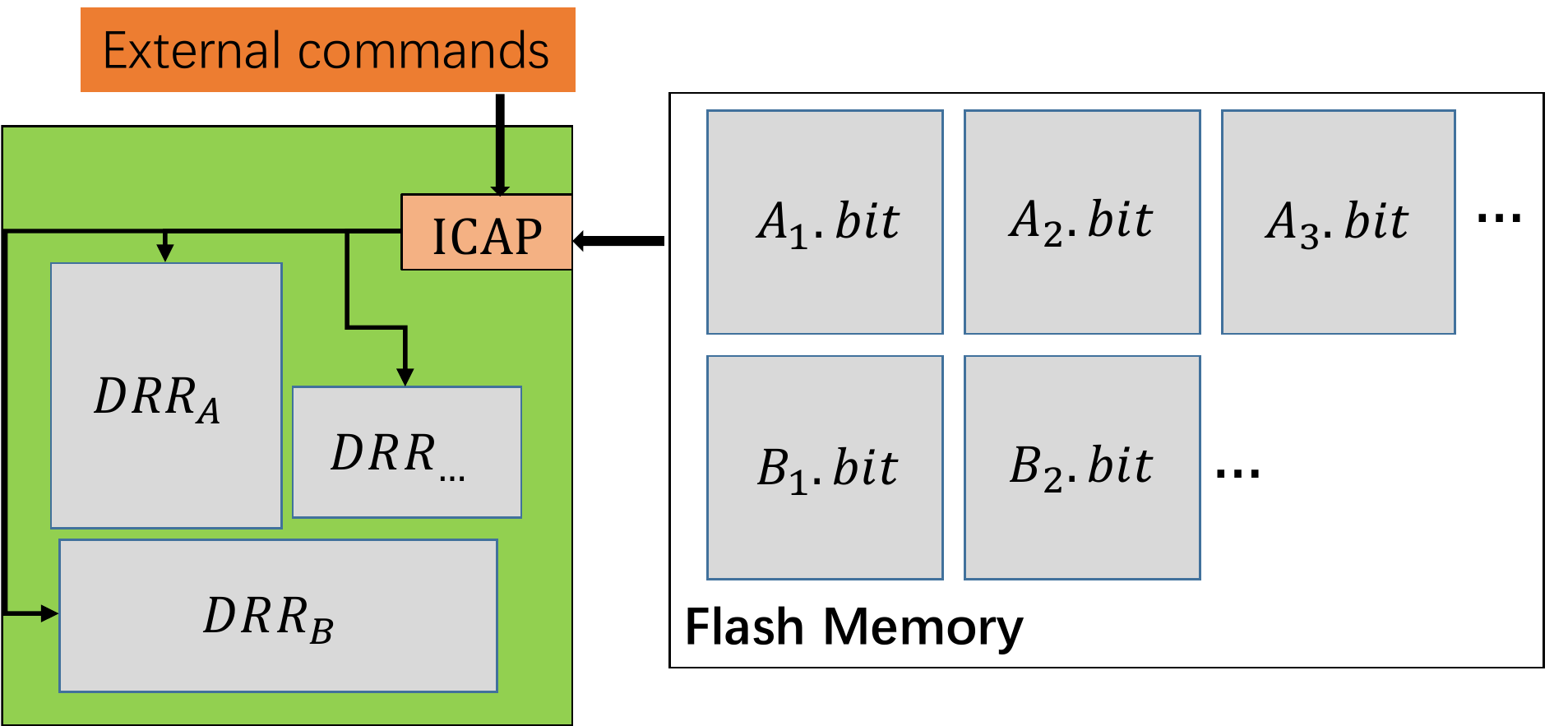}
	\caption{Partial dynamic reconfigurable process based on FPGA.}
	\label{fig:FPGA_PDRS}
\end{figure}

  	As shown in Fig. \ref{fig:FPGA_PDRS}, there are reconfigurable regions $DRR_A$, $DRR_B$ and so on.  
  	For  reconfigurable regions, there are corresponding configuration files, respectively.
  	For example, reconfigurable region $DRR_A$ has configuration files  $A_1.bit$, $A_2.bit$ and so on, while   reconfigurable region $DRR_B$ has configuration files $B_1.bit$, $B_2.bit$ and so on. 
  	These configuration files are stored in external  memory and will be transferred on chip at different times.  
  	When  the  commands of configuration control are running, these configuration files will be transferred on chip by the ICAP interface \cite{xilinx_pr}. Then  corresponding dynamic reconfigurable regions will be configured into corresponding circuits.  
  	As for different reconfigurable regions with corresponding configuration circuits at different times, we have the following definitions:
  
\begin{mydefinition}
	\label{DRR}
	${DRR} = \{drr_{i}|1 \leq i \leq N_{DRR} \}$ , $N_{DRR}$ is the number of reconfigurable regions. 
\end{mydefinition}
\begin{mydefinition}
	\label{TL}
	${TL} = \{(tl_{i}^{j}|1 \leq i \leq N_{DRR},1 \leq j \leq N_i^{tl}\} $,  $N_i^{tl}$ is the number of time layers in reconfigurable region $drr_{i}$.
\end{mydefinition}

\subsection{\textbf{Problem definition}}
\label{sec:problem_definition}
	To  deploy a large-scale computing application to a partially dynamically reconfigurable system based on FPGA, we need to divide the application into several task modules and then  partition, schedule and floorplan these task modules. 
	We define the set of $n$ task modules as $M=\{m_{i}|1 \leq i \leq n\}$.
	
	Implementing the reconfiguration process requires synthesizing task modules. The synthesizing results will provide the resources requirements of task modules, which mainly include  the demand for Lookup Table(LUT), Flip-Flop(FF), BRAM, and DSP.  While CLB is composed of LUT and FF, we can regard the synthesizing results as the demand for CLB, BRAM, and DSP.   
	The synthesizing and simulation results will also provide the execution time of each task module. In addition, to model the deployment of task modules in FPGA-PDRS, we add configuration time and shape to the task module. These notations and meaning related to the factors  of task module and the chip are shown in Table \ref{table:notations}.
\begin{table}[htbp]
	\centering
	\caption{Notations and meaning related to the factors of task module and chip}  
	\label{table:notations} 
	\begin{tabular}{l|l}  
		\hline 
		Notation&Meaning\\
		\hline
		$m_i$&task module i \\
		\hline
		$CLB_i$&task module $m_i$'s resource requirement for CLB  \\
		\hline
		$BRAM_i$&task module $m_i$'s resource requirement for BRAM  \\
		\hline
		$DSP_i$&task module $m_i$'s resource requirement for DSP \\
		\hline
		$width_i$&the width of task module $m_i$ \\
		\hline
		$height_i$&the height of task module $m_i$ \\
		\hline
		$exec_i$&the execution time of task module $m_i$ \\
		\hline
		$conf_i$&the configuration time of task module $m_i$ \\
		\hline
		$Width$ & the column number of the chip\\
		\hline
		$Height$& the row number of the chip\\
		\hline
	\end{tabular}
\end{table}	
	
	There are data dependencies among these task modules, which are defined by a task graph $TG = (V_{TG},E_{TG})$
	$V_{TG}$ represents the set of task modules $(=M)$.  $E_{TG}$ represents  data dependencies among  task modules.
	The definition of $E_{TG}$ is as follows:
\begin{mydefinition}
	\label{edge}	
	 $E_{TG} = \{edge_{i,j}^k = (m_{i},m_{j})^k|1 \leq i,j \leq n, i \ne j, k \leq N_{edge} \}$. $N_{edge}$ means the number of data dependencies among task modules; $edge_{i,j}^k$ means data produced by  $m_{i}$ will be used by $ m_{j}$ and is marked as the k-th data dependency. 
\end{mydefinition}	

	To ensure that the floorplan on heterogeneous resources is feasible, we consider the following constraints:
\begin{enumerate}
	\item The region occupied by the task module on the chip must meet the resources requirements of CLB, BRAM, and DSP.
	\item The task modules in the same time layer must be nonoverlapped and placed within their corresponding reconfigurable region. And the reconfigurable regions must be placed without overlapping each other.
	\item The whole part of the PDRS needs to meet the boundary constraints,  the width and the height of PDRS cannot exceed the width and the height of the chip, respectively.
	
\end{enumerate}
In summary, we define the input, output, and objective of this study as follows: 

\textbf{Input:} The graph of data dependency among task modules and related information of task modules.

\textbf{Output:} The solution of task modules partitioning, scheduling, and floorplanning in FPGA-PDRS.

\textbf{Objective:} Maximize the performance with improving the utilization of heterogeneous resources and reducing the communication cost. 

\subsection{ \textbf{Analysis of problems}}
\label{sec:problem_analy}

\subsubsection{Analysis of task modules partitioning, scheduling and floorplanning}
\label{sec:PSF}
	
	Partition   involves determining which task modules are in the same time layer of a reconfigurable region. 
	Schedule is to determine the configuration order of time layers with the task modules inside. 
	Floorplan is to determine the coordinates of reconfigurable regions and task modules. 
	
\begin{figure}[htbp]
		\centering
		\includegraphics[width=0.55\textwidth]{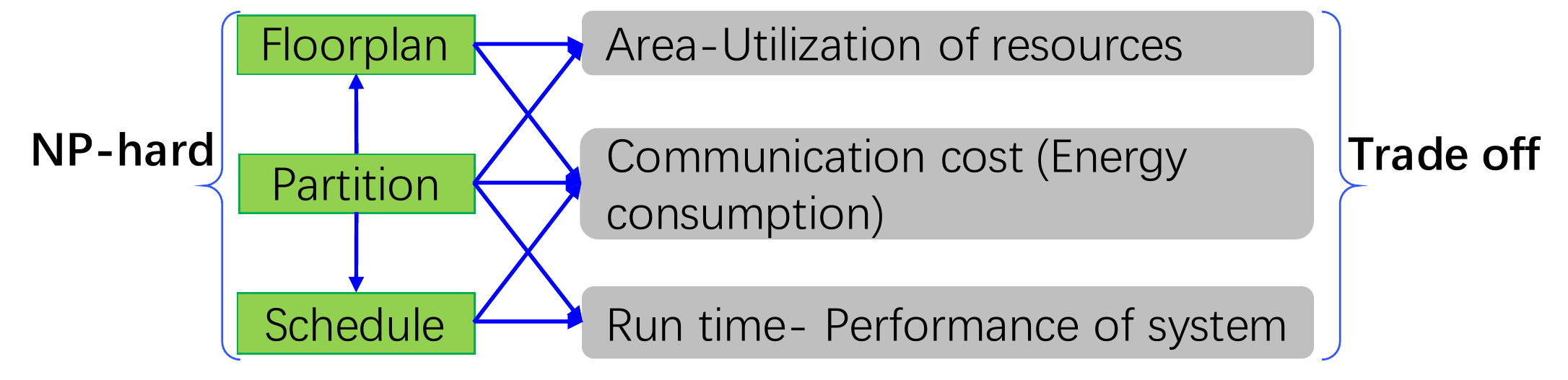}
		\caption{Partition, Schedule and Floorplan problems are NP-hard, affect each other and PDRS.}
		\label{fig:nphard}
	\end{figure}

	As shown in Fig. \ref{fig:nphard}, these three problems are NP-hard\cite{murata2003rectangle}\cite{hartmanis1982computers}.  
	They affect each other and affect the area, communication cost and performance of PDRS.   Task modules partitioning, scheduling, and floorplanning need to search for a proper solution that makes a trade-off between  the factors such as area, performance, and energy consumption.
	
\subsubsection{\textbf{The shape of task module}}
\label{sec:shape_modules}	
	 
	As described in Section \ref{sec:resouces_structure}, the on-chip programmable resources of FPGA are the column structure of CLB, BRAM and DSP. 
	And many previous works related to partition, schedule and floorplan are based on the given shape or the given area of task modules or reconfigurable regions. 
	However, after synthesizing each task module  by a logic synthesis tool such as Vivado ~\cite{Viavdo},   the synthesizing results we  obtain are the resources requirements of each task module. 	 	
	Moreover, the column structure of the FPGA chip is irregular.  
	The shapes of task modules mean the utilization of heterogeneous resources and are related to the location and resources requirements.
	For example,  assuming there is a task module  $m_1$. Its resources requirements for CLB, BRAM, and DSP are 20, 4, and 0, respectively.   
	As shown in Fig. \ref{fig:FPGA_resources}, $shape_1$ and $shape_2$, both can meet the resources requirements of task module $m_1$, are placed from coordinates $(4,0)$  and $(6,0)$, respectively.   
	The shape of $shape_1$ is $5*10$ with the utilization of resources  CLB, BRAM and DSP are $20/40, 4/4, 0$, respectively. While the shape of $shape_2$ is $8*5$ with the utilization of resources  CLB, BRAM and DSP are $20/30, 4/4, 0$, respectively.
	It is obvious that $shape_2$ is better than $shape_1$ if only in terms of resources utilization or area.
	
	There is no direct relationship between resources requirements and the shape of task module, and resource utilization can only be determined after determining the shape and the position of task module.
	The difficulty of this problem is to find the appropriate shapes and position  according to each task modules' resources requirements to ensure heterogeneous resources utilization.

\subsubsection{\textbf{The relationship between the shape of task modules and the shape of PDRS}}
\label{sec:shape_drr}		
\begin{figure}[htbp]
	\centering
	\includegraphics[width=0.43\textwidth]{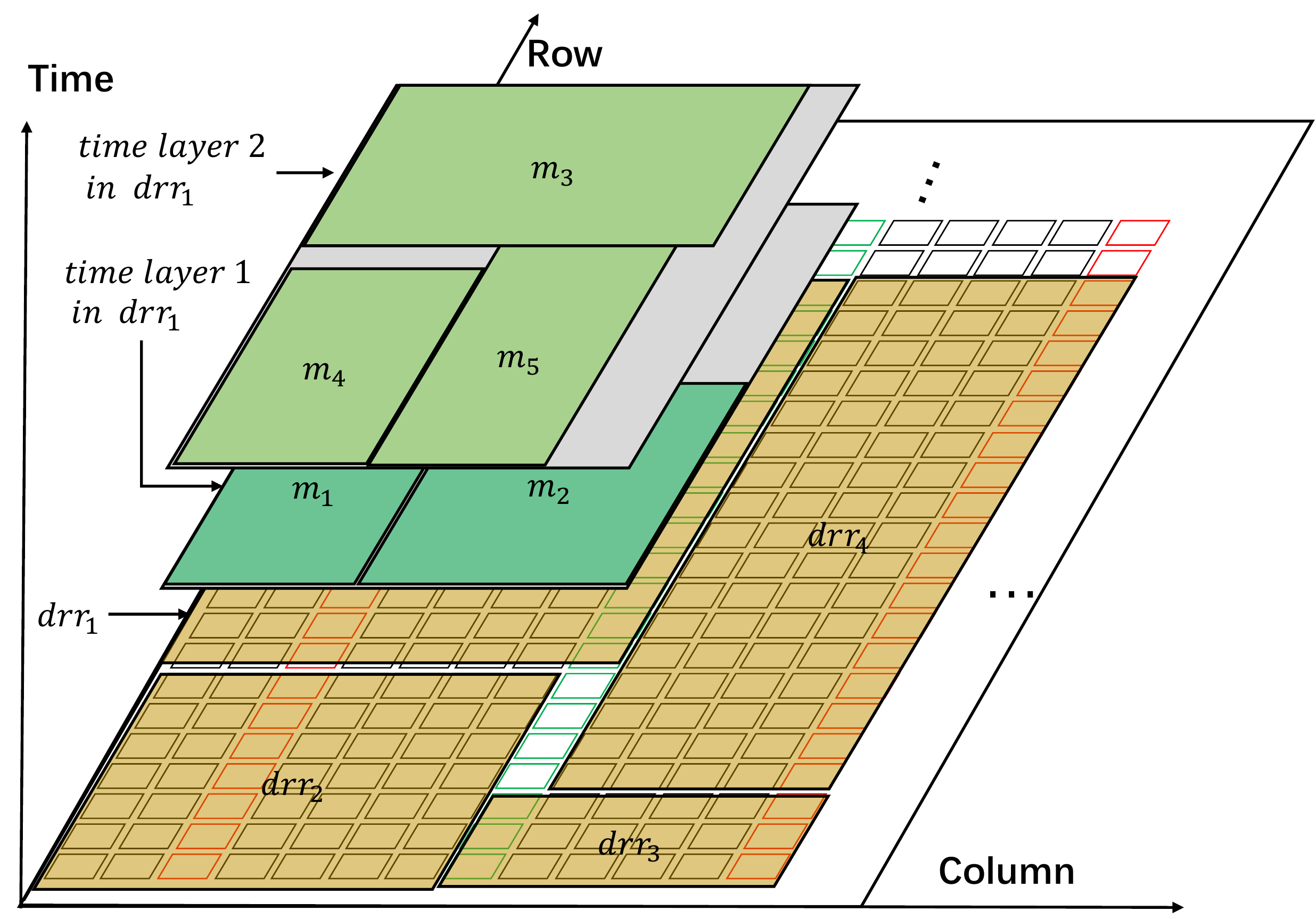}
	\caption{Different time layers in the same drr with different utilization of heterogeneous resources.}
	\label{fig:drr_shape}
\end{figure}

	As described in the Section \ref{sec:process_PDR}, on FPGA chip with PDR technology, the on-chip region is divided into one or several dynamically reconfgurable region(s).
	Moreover, the dynamically reconfigurable region is rectangular on FPGA chip, which is as shown  in Fig. \ref{fig:drr_shape}.
	Each dynamically reconfgurable region may have one or several time layer(s). Diﬀerent time layers implement diﬀerent functionalities and execute at diﬀerent times. And each time layer may have one or several task module(s). 
	
	There is no direct relationship between  shape of task modules and the shape  of PDRS. 
	In general, the shape of the task module affects the shape of its time layer, and the shape of this time layer affects the shape of the reconfigurable region that it belongs. Similarly, the shapes of the reconfigurable regions affect the shape of the whole PDRS.
	For example, as shown in Fig. \ref{fig:drr_shape}, dynamically reconfigurable region $drr_3$ has two time layers. Time layer $tl_3^1$ has two task modules $m_1$ and $m_2$ while time layer $tl_3^2$ has three task modules $m_3$, $m_4$ and $m_5$.   
	The shapes of time layers $tl^1_3$ and $tl^2_3$ are affected by the shapes of task modules that belong to them, respectively. The shapes of time layers $tl^1_3$ and $tl^2_3$ also affect the shape of dynamically reconfigurable region $drr_3$ and thus  affecting the shape of the whole PDRS. The difficulty of this problem is that there is no direct relationship between the shapes of task modules and the shape of PDRS. 

\section{Overall Design Flow}
\label{sec:representation-workflow}

\begin{figure}[htbp]
		\centering
		\includegraphics[width=0.7\textwidth]{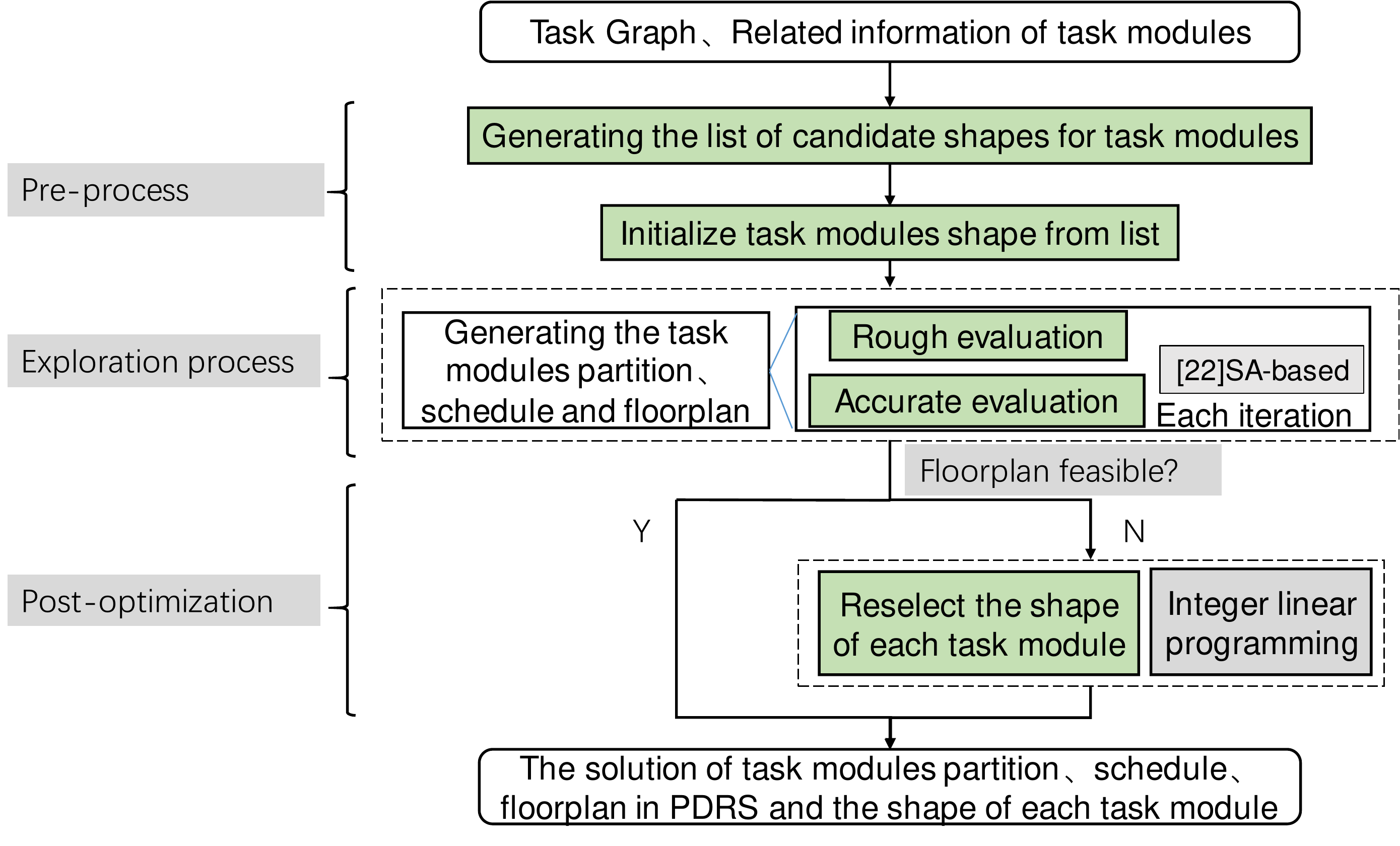}
		\caption{The complete workflow of generating task modules partition, schedule, floorplan on heterogeneous resources based on FPGA-PDRS.}
		\label{fig:overall_workflow}
	\end{figure}

	Fig. \ref{fig:overall_workflow} shows the complete workflow, and the colored parts are the contributions of this work. 	
	This workflow includes three parts, pre-processing,  the exploration of task modules partitioning, scheduling, floorplanning, and post-optimization. 
	
	The pre-processing involves generating the corresponding shape of the task module according to its resource requirements for CLB, BRAM, and DSP.  Due to the irregular column structure of heterogeneous resources on the chip,
	the generated shape needs to meet the resource requirements of the task module at each position on the chip. 	
	Different shapes mean different utilization of heterogeneous resources and affect the quality of the solution.
	Therefore, in this process, we generate a candidate shape list of each task module for the process of solution space exploration and post-optimization.
	Each shape in this shape list has a different width with corresponding height and  can meet the resource requirements of the task module at each position on the chip. This is illustrated in Section \ref{sec:shape-list} in detail. And then, we initialize a temporary shape for each task module from its candidate list by minimum area as the input of exploration process.

	In the process of exploring solution space, we change our previous work \cite{chen2018integrated}, which proposed the partitioned sequence triple to express  partition, schedule, and floorplan of task modules and generate the optimized solution of PDRS  but modeling homogeneous resource with simplifying these problems.  
	This work is SA-based, and only considers the area cost, scheduling cost, and communication cost.
	
	The exploring process use the partitioned sequence triple P-ST(PS, QS, RS), which was proposed by \cite{chen2018integrated} to represent the partition, schedule, and the floorplan of task modules in PDRS. In P-ST (PS,QS,RS), (PS,QS) is a hybrid nested sequence pair (HNSP) and represent the spatial and temporal partitions, as well as the ﬂoorplan,  RS is the partitioned dynamic configuration order of the time layer with its task modules.
	
	In the exploring process, we first generate an initial solution and then search the optimal solution through the SA-based search engine. In each iteration process of the SA-based algorithm, we delete a task module from  P-ST and then find a new location for insertion to generate a new solution. 
	In the process of finding the optimal insertion location, there are two parts: rough evaluation and accurate evaluation. The rough evaluation evaluates the cost of scheduling time and area in linear time for all feasible insertion points, and then accurate evaluation will evaluate the selected insert locations from rough evaluation by recalculating the cost to find the best solution for this disturbance. 
	
	In the rough evaluation phase, when we evaluate the area cost of each insert point, we evaluate all the shapes in the candidate shape list of this deleted task module for the current insert point. 
	And we choose the shape of this deleted task module that minimizes the area of the whole design of PDRS to evaluate this insert point.  
	In this way, we explore the possibility of more floorplan solutions. And because the task module shape may be changed, the area cost may be reduced, thus making the solution with greater probability tends to optimize the scheduling time and communication cost.
	
	In the accurate evaluation phase, we calculate the resource usage of the reconfigurable regions and calculate the cost of  heterogeneous resource utilization and add this cost to the total cost. The cost of heterogeneous resource utilization is calculated by Equation (\ref{eq:cost_hetero_accurate}). In Equation (\ref{eq:cost_hetero_accurate}), $CLB_{have}$, $BRAM_{have}$, and $DSP_{have}$ mean the resources that the chip has, $CLB_{use}$, $BRAM_{use}$, and $DSP_{use}$ mean the resources that reconfigurable regions use. Then the total cost function is shown by Equation (\ref{eq:cost}), and all cost values are normalized.
	
	In Equation (\ref{eq:cost}),  the objection function $Cost_{total}$ is defined as a linear combination of schedule length, area cost, communication cost, and the cost of heterogeneous resources utilization. The area cost $cost_{area}$ includes the cost of exceeding  the boundary of the chip.  Decreasing this cost means trying to make the floorplan of PDRS can meet the boundary constraint. The cost $cost_{schedule}$ is the schedule time of PDRS, which means the performance and is the critical path from the beginning of the configuration process
	to the end of the execution of all task modules.   The cost $cost_{comm}$ is the communication cost of the PDRS, which means the cost of data transfer. And the cost $cost_{hetero}$ means the utilization of heterogeneous resources, decreasing this cost means trying use more heterogeneous resources on the chip  and thus improving the performance and reducing the communication cost of PDRS.

\begin{equation}
	\label{eq:cost}
	\begin{aligned}
		Cost_{total} = & \alpha \times cost_{area} + \beta \times cost_{schedule} + \gamma \times cost_{comm}+ \lambda \times cost_{hetero}
	\end{aligned}
\end{equation}

\begin{equation}
	\label{eq:cost_hetero_accurate}
	\begin{aligned}
		cost_{hetero} = \frac{CLB_{have}}{CLB_{use}} + \frac{BRAM_{have}}{BRAM_{use}} + \frac{DSP_{have}}{DSP_{use}}
	\end{aligned}
\end{equation}
	After several iterations of the SA-based search engine, the framework generates the solution of task modules partitioning, scheduling, and ﬂoorplanning for the PDRS.
	The solution means that task modules' partition, schedule, and floorplan of PDRS are determined. As shown in Fig. \ref{fig:drr_shape}, each task module's time layer and each time layer's reconfigurable region are determined. The relative position relationships among task modules and the relative position relationships among reconfigurable regions are also determined.

	To better represent the results of partition, schedule, and floorplan of task modules, we have the following definitions. 
	
	\begin{mydefinition}
		\label{module_drr}
		If task module $m_i$ is partitioned into reconfigurable region $drr_j$, then $drr(m_i) = drr_j$ .
	\end{mydefinition}
	\begin{mydefinition}
		\label{module_tl}
		If task module $m_i$ is partitioned into the k-th time layer of reconfigurable region  $drr_j$, then $tl(m_i) = tl_j^k$ .
	\end{mydefinition}
	
	And we can use the Definition. \ref{module_drr} and Definition. \ref{module_tl} to express the   relationship of partition between task modules.
	If $drr(m_i) == drr(m_j)$, task modules $m_i$ and  $m_j$ are partitioned into the same reconfigurable region. And in contrast, if $drr(m_i) \neq drr(m_j)$,  the two task modules are in different reconfigurable regions.
	If $tl(m_i) == tl(m_j)$, task modules $m_i$ and  $m_j$ are partitioned into the same time layer of a reconfigurable region. 
	If $drr(m_i) == drr(m_j)$ and $tl(m_i) \neq tl(m_j)$, task modules $m_i$ and  $m_j$ are partitioned into the same reconfigurable region but different time layers.
	
	In the process of exploring the solution space,  there are the costs of schedule length,  area, communication, and the utilization of heterogeneous resources. The process of  exploring the solution space will make a trade-off among these costs and may generate a solution that the width or height of PDRS exceeds the boundary of the chip, which means the solution is infeasible.  
	
	If this solution is not feasible due to boundary constraint, we can try to make it meet the solution by the process of post-optimization. Based on these information of task modules' partition, schedule, and floorplan and the candidate shape list that are generated in the process of pre-processing, we can reselect the shapes for all task modules to make the floorplan of PDRS meet the boundary constraint and thus improving the success rate of floorplan.
	In the process of post-optimization, we build an ILP model to reselect the shapes of task modules to improve the floorplan success rate of PDRS.
	The ILP model constrains the shape of task modules and the position relationship among task modules under the given solution of task module partition, schedule, and floorplan.  Then we add the boundary constraints to this model and set the objective function of the model. Finally, we use the solver Gurobi\cite{gurobi} to solve this model. The ILP model is illustrated in Section \ref{sec:select-shape} in detail.
	
	In this complete workflow, the pre-processing process, solution space exploration process, and post-optimization process are three relatively independent processes. These three processes affect the utilization of heterogeneous resources, the quality of the solution, and the floorplan success rate of the solution.  
	After these three processes, we get the shape of each task module and the optimized solution for task module partitioning, scheduling,  and floorplanning of PDRS.

\section{Proposed Algorithms}\label{sec:Algorithms}
\subsection{Generation of Task Modules Shape List}\label{sec:shape-list}
Each task module has the demands for heterogeneous resources CLB, BRAM, and DSP on the chip. For the task module $m_i$, the demand is denoted as $[CLB_i,BRAM_i,DSP_i]$.   
The irregular column structure on the chip and undetermined position of task modules make generating the width and height of task module $m_i$, which are denoted as $width_i$ and $height_i$, respectively, a tough problem. 

The shape of the task module $m_i$ is larger will reduce heterogeneous resources utilization. In contrast, the smaller shape of the task module $m_i$   may  not meet the resource requirements in a certain position, thus inducing infeasible floorplan.  
To make the shape of task module $m_i$  meet the resource requirements at each position on the chip, the shape needs to move along the x-axis and meet the resource requirements in each movement.  


\begin{algorithm}[htbp]
	\caption{Generation of task modules shape List}
	\label{alg1}
	\begin{algorithmic}
		\State{\textbf{Input}:Module resources requirements, and the irregular column structure of chip.}
		\State{\textbf{Output}:The candidate list with $N$ candidate shapes of each task module.}
		\State{ M $\gets$  task module set; L $\gets$ $\emptyset$.}   \Comment{L : The set of candidate shape list}
		\For{$m_i$ in  M}	
		\State{ $width_{init}$  $\gets$ $\max\{ CLB_{i}/Height_{CLB}, BRAM_{i}/Height_{BRAM},DSP_{i}/Height_{DSP} \}$}
		\State{$N_{shape}^i$ $\gets$ Width - $width_{init}$}  \Comment{$N_{shape}^i$: the shape number of $m_i$; Width : The width of chip}
		\For{j $\gets$ 0 to $N_{shape}^{i}$}
		\State{$width_i^j$ $\gets$ $width_{init}$}
		\State{$height_i^j$ $\gets$ Height}  \Comment{Height : The height of chip}
		\State{$Flag_{move}$  $\gets$ Check feasible in each position of the chip}
		\If{$Flag_{move}$ == 0}
		\State{Continue}
		\Else
		\While{$Flag_{move}$ == 1}
		\State{$height_i^j$ $\gets$ $height_i^j$ - 1}
		\State{$Flag_{move}$  $\gets$ Check feasible in each position of the chip}
		\If{$Flag_{move}$ == 0}
		\State{$height_i^j$ $\gets$ $height_i^j$ + 1}
		\State{Break}
		\EndIf
		\EndWhile
		\EndIf	
		\State{$shape_i^{j}$ $\gets$ $[width_i^j,height_i^j]$}
		\If{aspect ratio of $shape_i^j$ $\leq$ $\gamma$}
		\State{$L_{i}.append(shape_i^j)$}
		\EndIf
		\State{$L_{i}$ $\gets$ remove shapes with the same height but larger width}
		\State{$L_{i}$ $\gets$ Select top $n$  shapes from $L_{i}$ according to area of shape}
		\EndFor
		\EndFor	
	\end{algorithmic}
\end{algorithm}

	
	Since the irregular column structure on the chip is continuous, we adopt an enumeration strategy. The pseudo code for generating the candidate shape list is shown in Algorithm. \ref{alg1}. We traverse all task modules and all the possible widths of each task module on the chip. And for each width of task module $m_i$, we find a height that can let the task module meet the resource requirements at all positions on the chip.	
	Specifically, under each width of task module $m_i$, if the maximum height can meet the resource requirements at each position on the chip, we reduce the height one by one until it cannot meet this condition.  
	Since we use the enumeration strategy of increasing the width and determining the minimum height under each width, some shapes with the same height but different widths may appear. 
	Those shapes with smaller widths are better shapes. Thus, we remove those shapes with the same height but larger width from the list.
	Then if the aspect ratio of the shape is less than $\gamma$, we add this shape to the candidate shape list.  
	Since \cite{chen2008fixed} interpret intuitively  and show by experiments  that  an aspect ratio close to one is  beneficial to wirelength, we set $\gamma$ to 1.5 for optimization of communication cost and ensure a certain solution space margin.
	
	Finally, according to area, we choose the top $n$ shapes from this list as the final candidate shape list.
	The value of $n$ is related to the complexity of the exploration process and the process of post-optimization. 
	A relatively larger value of $n$ means that the exploration process can explore more floorplan solutions and better optimize schedule time and communication cost.  Also, the larger value of $n$  provides a  larger solution space of reselecting task modules shapes for the process of post-optimization and can better improve the floorplan success rate of the solution.	  And in this work, we empirically set $n$ to 10. 

	After generating the candidate shape list for each task module, we initialize the shape of each task module from the corresponding list. In this paper, we use the shape with minimum area as the temporary shape of the task module.
	Then we  use these initialized task modules shape as the input of the exploration process to generate the solution of task modules partitioning, scheduling, and floorplanning of PDRS.

\subsection{Reselection of Task Modules Shape}
\label{sec:select-shape}

\begin{figure}[htbp]
	\centering
	\subfloat[]{\includegraphics[width=0.4\textwidth]{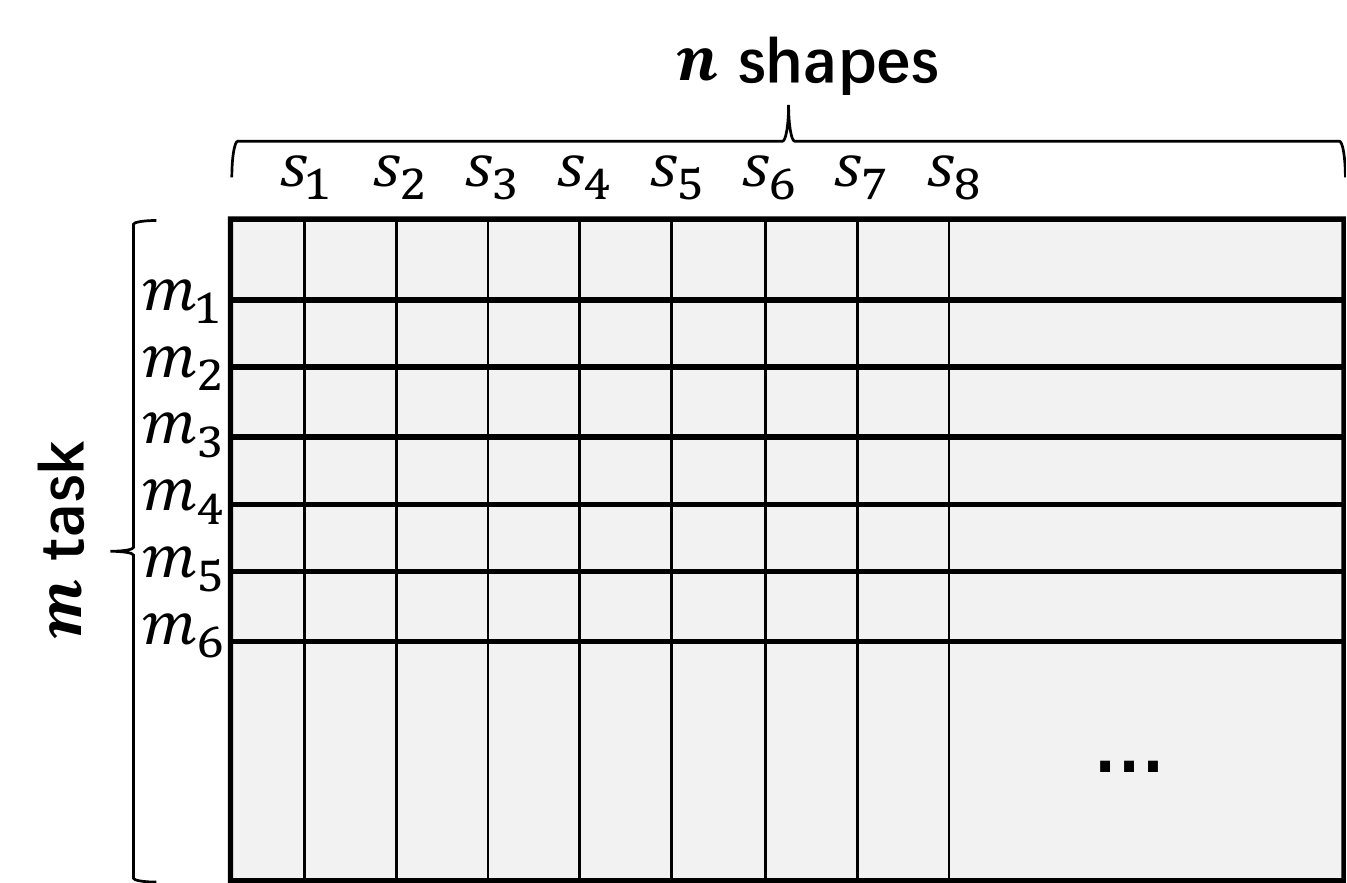}\label{fig:select_shape_matrix}}
	\hspace{.5in}
	\subfloat[]{\includegraphics[width=0.5\textwidth]{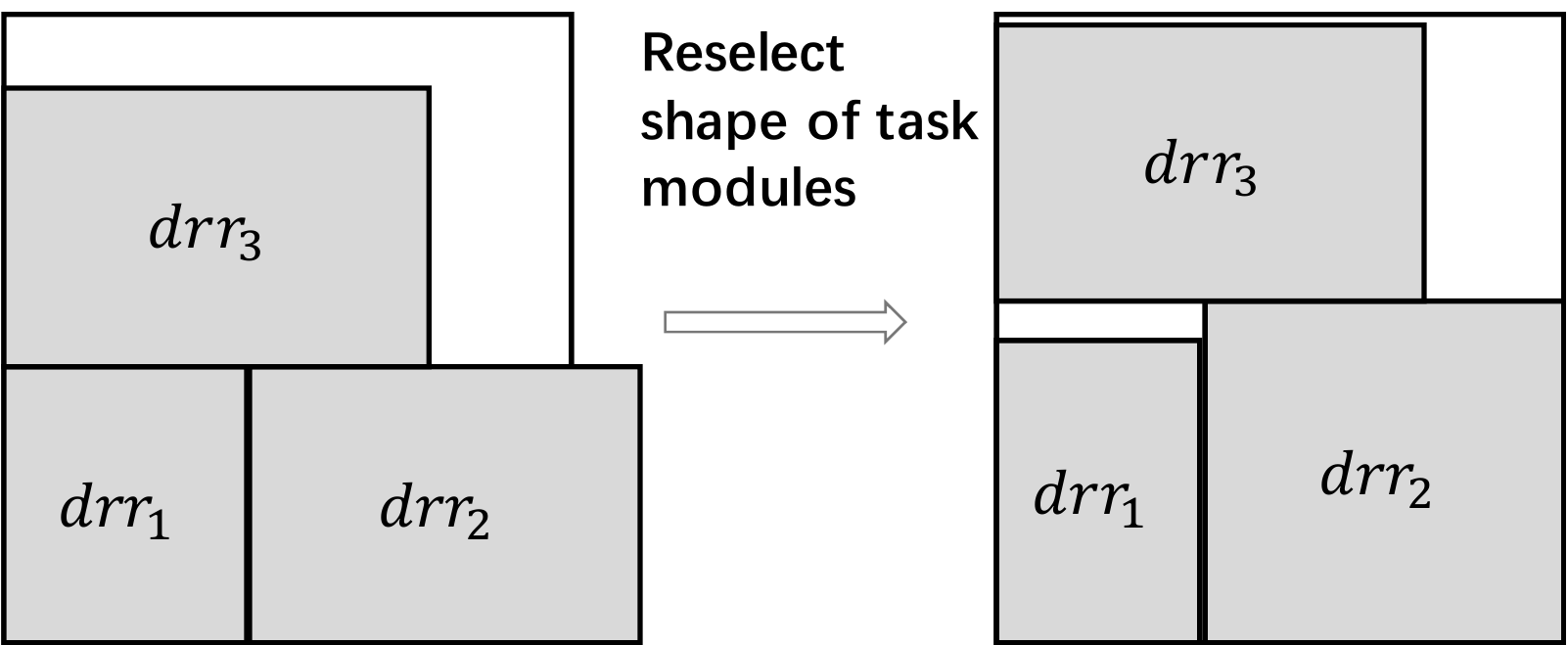}\label{fig:reselect_shape_exam}}
	\caption{(a)The matrix of candidate shape list for task modules. (b) The example of reselect shape to meet boundary constraint.}
	
\end{figure}

	 The process of exploring the solution space reduces the total cost in each iteration. And to ensure the performance of PDRS, the floorplan of the generated solution may exceed the boundary, which is like the left figure of Fig. \ref{fig:reselect_shape_exam}.   As described in Section \ref{sec:shape_drr}, there are no direct relationships  between the task module shapes and the shape of the whole PDRS,   and  the shape of the task module affects the shape of its time layer and thus affects the shape of the reconfigurable region.  To make the floorplan of this solution feasible, we can reselect the task modules' shapes from the candidate shape lists, and then change the shape of reconfigurable regions thus generating the feasible floorplan like the right figure of  Fig. \ref{fig:reselect_shape_exam}.
		
	As shown in Fig. \ref{fig:select_shape_matrix}, assuming there are $m$ task modules, and there are $n$ shapes in the list of each task module. Then the solution space size is $n^m$. 
	Therefore we establish a ILP model to solve it and try to find feasible floorplan of the solution through the solver Gurobi. The ILP model involves three-part constraints, one is the constraint of task modules shape, one is the constraint of the position relationship among task modules, and one is the boundary constraint of the whole design of PDRS.

	We use a matrix to represent the shape selection of task modules, which likes Fig. \ref{fig:select_shape_matrix} and denoted as $MS$. 
	If task module $m_i$ use shape $s_j$, then we denote this by $MS[i,j] = 1$. 
	A task module can only have one shape, so we have the following constraint for selecting the shape of each task module.  This constraint is shown by Equation (\ref{eq:select_shape}), and in  Equation (\ref{eq:select_shape}), $m$ is the number of task modules.
	\begin{equation}
		\label{eq:select_shape}
		\forall{m_{i}} \in M, 1 \leq i \leq m, 
		\sum_{s_j,j=1}^{n}{MS[i,j]} = 1 
	\end{equation}

	Then, the width and height of task module $m_i$ can be expresented by linear expression Equation (\ref{eq:module_width}) and Equation (\ref{eq:module_height}), respectively.
	\begin{equation}
		\label{eq:module_width}
		width_i = \sum_{s_j,j=1}^{n}{MS[i,j] \times width_i^j}
	\end{equation}
	\begin{equation}
		\label{eq:module_height}
		height_i = \sum_{s_j,j=1}^{n}{MS[i,j] \times height_i^j}
	\end{equation}

	After adding constraints to the shape of task modules, we then add constraints to the positional relationship among task modules and finally get the boundary of the whole design of PDRS.
	In this work, we use \textit{P-ST} that is proposed in our previous work \cite{chen2018integrated} to represent the solution information. 
	\textit{P-ST} includes the Sequence Pair(SP)\cite{murata1996vlsi} to represent floorplan, and the information of reconfigurable regions and time layers are added to \textit{SP} to express partition.

\begin{algorithm}[htbp]
	\caption{Constraint of position relations in ILP model}
	\label{alg:qp}
	\begin{algorithmic}
		\State{$iter \gets  0$}  
		\For{$m_i$ in $ms$}
		\State{Constraint: $Mat_x[iter][m_i]$ $\geq$ $Mat_x[iter][m_{i-1}] + width_i$ }  \Comment{Update linear expressions of horizontal position}
		\State{$iter \gets iter + 1$} 
		\For{$m_j$ in $ps[m_i :]$}
		\If {$drr(m_i)$ != $drr(m_j)$  or $tl(m_i)$ == $tl(m_j)$}
		\State{Constraint: $Mat_x[iter][m_j]$ $\geq$ $Mat_x[iter][m_i])$} \Comment{Constrained coordinate of horizontal position relation}
		\EndIf
		\EndFor
		\For{$m_j$ in $ps$}
		\If {$drr(m_i)$ != $drr(m_j)$  or $tl(m_i)$ == $tl(m_j)$}
		\State{Constraint: $Mat_x[iter][m_j]$ $\geq$ $Mat_x[iter-1][m_j])$ } \Comment{Store linear expressions to update}
		\EndIf
		\EndFor	
		\State{Constraint: $Mat_y[iter][m_i]$ $\geq$ $Mat_y[iter][m_{i+1}] + height_i$ }   \Comment{Update linear expressions of vertical position}
		\For{$m_j$ in $ps[: m_i]$}
		\If {$drr(m_i)$ != $drr(m_j)$  or $tl(m_i)$ == $tl(m_j)$}
		\State{Constraint: $Mat_y[iter][m_j]$ $\geq$ $Mat_y[iter][m_i])$}   \Comment{Constrained coordinate of vertical position relation}
		\EndIf
		\EndFor
		\For{$m_j$ in $ps$}
		\If {$drr(m_i)$ != $drr(m_j)$  or $tl(m_i)$ == $tl(m_j)$}
		\State{Constraint: $Mat_y[iter][m_j]$ $\geq$ $Mat_y[iter-1][m_j])$} \Comment{Store linear expressions to update}
		\EndIf
		\EndFor
		\EndFor	
	\end{algorithmic}
\end{algorithm}

	All constraints of calculating coordinates are added to these two matrices. By traversing \textit{SP} and adding constraints to each row of the two matrices,
	we can get the linear expression of the constraints coordinate of the task modules floorplan.  And for the boundary of  floorplan of PDRS, we have the following constraints Equation (\ref{eq:x_max}) and Equation (\ref{eq:y_max}) to limit $X_{max}$ and $Y_{max}$, respectively. 
	\begin{equation}
		\label{eq:x_max}
		\forall{value_x} \in Mat_x[m], value_x \leq X_{max}, 
	\end{equation}
	
	\begin{equation}
		\label{eq:y_max}
		\forall{value_y} \in Mat_y[m], value_y \leq Y_{max},
	\end{equation}

	The feasible floorplan means that the whole PDRS should be placed within the chip boundary, therefore we add the constraints to the boundary of PDRS. The constraints are shown by Equation (\ref{eq:xy_max_cons}), and in Equation (\ref{eq:xy_max_cons}), $Width$ and $Height$ are the boundaries of the chip.
	\begin{equation}
		\label{eq:xy_max_cons}
		  X_{max} \leq Width , Y_{max} \leq Height
	\end{equation}
	
	Finally, we set the objective function of this model as to maximize the value of subtracting between chip boundary and the boundary of the whole design PDRS. This objective is shown by Equation (\ref{eq:object}).
	

\begin{equation}
	\label{eq:object}
	\max\{ Width - X_{max} + Height-Y_{max}\} 
\end{equation}

\section{Experiments and Results}\label{sec:experiments}

\subsection{Experimental Setup}

	In this study, all proposed methods  were implemented in C/C++ language on a 64-bit Linux  workstation (Intel(R) Xeon(R) Gold 6254 CPU @ 3.10 GHz, 125 GB RAM).
	Since there are no standard benchmarks for the discussed problems, we constructed the benchmarks by using the tool Task Graphs For Free (TGFF) ~\cite{1998TGFF} to generate a directed acyclic graph(DAG) to represent the task modules and the data dependencies among task modules. 	We denote this graph as task graph(TG).
	The information of nodes and edges, which represent task modules and data dependencies, respectively, in TG, is shown in Table \ref{tab:benchmark}.   
	
	We also generate two benchmarks from two popular convolutional neural network models, AlexNet\cite{krizhevsky2017imagenet} and VGG\cite{simonyan2014very}. The two benchmarks respectively include all the convolution layers, pooling layers, and full connection layers of the two neural networks. 
	The convolution layer and full connection layer mainly include the multiply-add computing unit, and the pooling layer mainly includes the comparator. 
	Therefore, we synthesize the heterogeneous resources usage of the multiply-add computing unit and the comparator, respectively.  We use several multiply-add computing units as a convolution kernel and several comparators as a pooling kernel. 
	For the convolution layer and full connection layer with larger computation,  we use several convolution cores to perform convolution operation over these feature maps by channel parallel, and for the pooling layer, we use a specified pooling core to perform pooling operations over all the output feature maps. The execution times of task modules are estimated based on a frequency of 200MHz.
\begin{table*}[htbp]
	\centering\small
	\caption{Benchmark Information}
	\label{tab:benchmark}
	\begin{tabular}{c|c|c|c|c|c|c|c|c|c}
		\hline
		\multirow{2}{*}{Bench.} & \multirow{2}{*}{imp.} & \multicolumn{8}{|c}{TG}\\
		\cline{3-10}
		& & \#V & \#E& VWR $(ms)$& EWR & CPT $(ms)$&CLB&BRAM &DSP\\
		\hline
		\multirow{3}{*}{t10}
		& 1	&	\multirow{3}{*}{10}  & 8   	& (40,55)    & (20,30)   & 151.1 	&(2000,3000)&(0,80) &(0,80)  \\
		& 2	&   					 & 10   & (40,55)    & (20,30)   & 259.2 	&(2500,3500)&(20,100) &(20,100)  \\
		& 3	&						 & 12   &(40,55)   	 & (20,30)   & 359.1 	&(3000,4000)&(40,120) &(40,120)    \\
		
		\hline
		\multirow{3}{*}{t30}	
		& 1	&\multirow{3}{*}{30} 	& 71   	& (40,60)    & (20,30)   & 727.6    &(2000,3000)&(0,80) &(0,80)  \\
		& 2	&    					& 51   	& (30,350)   & (60,610)  & 1450.8 	&(2500,3500)&(20,100) &(20,100)  \\
		& 3	& 					    & 72   	& (40,60)    & (20,30)   & 786.3 	&(3000,4000)&(40,120) &(40,120)  \\
		
		\hline
		\multirow{3}{*}{t50}
		& 1	&\multirow{3}{*}{50}	& 78    & (40,60)    & (20,30)   & 595.9 	&(2000,3000)&(0,80) &(0,80)   \\
		& 2	&    					& 33    & (40,60)    & (20,30)   & 244.0 	&(2500,3500)&(20,100) &(20,100)  \\
		& 3	&						& 51    & (20,180)   & (50,350)  & 464.9 	&(3000,4000)&(40,120) &(40,120)   \\
		
		\hline
		\multirow{3}{*}{t100}
		& 1	& \multirow{3}{*}{100}  & 110   & (20,180)   & (50,350)  & 703.1 	&(2000,3000)&(0,80) &(0,80) \\
		& 2	&     					& 134	& (20,180) 	 & (50,350)  & 1265.6 	&(2500,3500)&(20,100) &(20,100)   \\
		& 3	&	  					& 147	& (20,180)	 & (50,350)  & 580.5    &(3000,4000)&(40,120) &(40,120)   \\
		\hline
		\multirow{3}{*}{t200}
		& 1	&	\multirow{3}{*}{200} & 403	& (10,390)	 & (30,770)  & 1482  	&(2000,3000)&(0,80) &(0,80)  \\
		& 2	&   					& 312   & (10,390)   & (30,770)  & 3556.2 	&(2500,3500)&(20,100) &(20,100)  \\
		& 3	&	  					& 327	& (40,60)	 & (20,30)   & 436.4  	&(3000,4000)&(40,120) &(40,120) \\		
		\hline
		\multicolumn{2}{c|}{AlexNet}	&17	& 26 &(0.7,7.5) &(4,189.1) 	&34.4 	&(400,1800)&(5,10) &(0,8)\\
		\hline
		\multicolumn{2}{c|}{VGG}		&47	&108 &(5.1,61.7)&(4,784) 	&498.2	&(400,1800)&(5,10) &(0,8)\\
		\hline

		
	\end{tabular}
\end{table*}

In Table \ref{tab:benchmark}, each benchmark has three different implementations and these implementations has the same number of task modules but different task dependencies.   And $\#V$ and $\#E$ denote the number of nodes and edges in TG, respectively. 
The column $VMR$ represents the range of random values of execution time for task modules, the column $EWR$ means the range of random values of communication among task modules, and the column $CPT$ represents the critical paths of the task graphs, in which the task modules are weighted by execution time. The column $CLB$, $BRAM$, and $DSP$ represent the range of random values of task modules' resources requirements for CLB, BRAM, and DSP.
And in this work, we  use one of the most widely used Xilinx Virtex-7  series FPGA chips, \textit{XC7VX485T}, as the target chip.  


\subsection{Experimental results and analysis}
\label{sec:self-comp}
	In this section, we show and analyze the experimental results of the comparison. 
	The complete workflow in this paper includes three parts, pre-processing,  the exploration of task modules partitioning, scheduling, floorplanning, and post-optimization.  
	To demonstrate the efficiency and effectiveness of the three processes, we accomplish three comparative experiments. 
	
	Firstly, we compare the effect of our method  of generating task modules shapes on heterogeneous resources usage  with  \cite{banerjee2009fast}. 
	Secondly, we compare the solution generated  by our complete workflow with the solution generated by the method  $Int\_PSF$ in previous work\cite{chen2018integrated} on schedule time, communication cost,  floorplan success rate, and heterogeneous resources reuse rate. 
	Thirdly, We compare the improvement of the floorplan success rate through the ILP model in post-optimization with the solution generated by the exploration process.
	
\subsubsection{\textbf{Comparison of the generated task module shapes}}
\label{sec:init_shape_task}
	The generation of task modules shape involves the area of task modules and the utilization of heterogeneous resources. 
	P. B et al.\cite{banerjee2009fast}  found the smallest width on the chip that includes CLB, BRAM and DSP resources and then generated the height of task modules based on this width. However, this method is not conducive to the utilization of resources. 
	To demonstrate the efficiency of our initialization strategy of task module shapes and show the impact  on task module area and resource utilization, we have compared these results with \cite{banerjee2009fast} and show the experiment results in Table \ref{tab:init_moduleshape_cmp}. 
	The columns $CLB$, $BRAM$, and $DSP$ in the column $Need\ Resources$ mean the number of resource requirements of CLB, BRAM, and DSP for each task module in the corresponding benchmark, respectively, on average.  
	In the columns $[34]$ and $Our\ method$, the $CLB$, $BRAM$, $DSP$,  mean the number of corresponding resources to be used on the chip of each task module, respectively, on average. And the column $Area$ means means the total area of the task module shapes in the corresponding benchmark. 

	The experimental results show that, compared with \cite{banerjee2009fast},   our task module shapes generation method can  reduce the occupied resources while meeting the resource requirements of each position on the chip. 
	And as the average occupied area is decreased by 6\%, the occupied resources of CLB, BRAM and DSP are decreased by 6.6\%, 13.2\%, and 4.7\%, respectively, on average. 
\begin{table}[htbp]
	\centering\small
	\caption{Comparison of the generated task module shapes}
	\begin{tabular}{|ll|lll|llll|llll|}
		\hline
		\multirow{2}{*}{Bench.} &
		\multicolumn{1}{c|}{\multirow{2}{*}{Imp.}} &
		\multicolumn{3}{c|}{Need resources} &
		\multicolumn{4}{c|}{{\cite{banerjee2009fast}}} &
		\multicolumn{4}{c|}{Our   method} \\ \cline{3-13} 
		&
		\multicolumn{1}{c|}{} &
		\multicolumn{1}{c}{CLB} &
		\multicolumn{1}{c}{BRAM} &
		\multicolumn{1}{c|}{DSP} &
		\multicolumn{1}{c}{CLB} &
		\multicolumn{1}{c}{BRAM} &
		\multicolumn{1}{c}{DSP} &
		\multicolumn{1}{c|}{Area} &
		\multicolumn{1}{c}{CLB} &
		\multicolumn{1}{c}{BRAM} &
		\multicolumn{1}{c}{DSP} &
		\multicolumn{1}{c|}{Area} \\ \cline{1-2}
		\hline
		\multirow{3}{*}{t10}  & 1 & 2388 & 33 & 36 & 2622 & 139 & 214 & 35321   & 2484 & 125 & 206 & 33507  \\
		& 2 & 2977 & 56 & 62 & 3282 & 174 & 269 & 44213   & 3080 & 154 & 258 & 41538  \\
		& 3 & 3437 & 76 & 86 & 3852 & 204 & 315 & 51889   & 3555 & 175 & 296 & 47966  \\
		\hline
		\multirow{3}{*}{t30}  & 1 & 2454 & 39 & 39 & 2713 & 144 & 222 & 109649  & 2550 & 128 & 212 & 103181 \\
		& 2 & 2981 & 63 & 56 & 3268 & 173 & 268 & 132069  & 3088 & 154 & 258 & 124949 \\
		& 3 & 3518 & 84 & 82 & 3927 & 208 & 321 & 158707  & 3637 & 179 & 304 & 147233 \\
		\hline
		\multirow{3}{*}{t50}  & 1 & 2569 & 36 & 42 & 2807 & 149 & 230 & 189088  & 2670 & 133 & 222 & 180042 \\
		& 2 & 2995 & 60 & 58 & 3291 & 175 & 269 & 221692  & 3108 & 155 & 259 & 209584 \\
		& 3 & 3540 & 79 & 76 & 3908 & 207 & 320 & 263226  & 3660 & 180 & 305 & 246903 \\
		\hline
		\multirow{3}{*}{t100} & 1 & 2455 & 39 & 38 & 2703 & 144 & 221 & 364078  & 2550 & 127 & 212 & 343912 \\
		& 2 & 3006 & 60 & 60 & 3316 & 176 & 272 & 446709  & 3118 & 155 & 260 & 420341 \\
		& 3 & 3482 & 75 & 76 & 3837 & 204 & 314 & 516952  & 3600 & 177 & 300 & 485864 \\
		\hline
		\multirow{3}{*}{t200} & 1 & 2491 & 41 & 41 & 2743 & 146 & 225 & 739157  & 2588 & 129 & 215 & 698020 \\
		& 2 & 2962 & 61 & 60 & 3284 & 174 & 269 & 884697  & 3073 & 153 & 257 & 828717 \\
		& 3 & 3511 & 83 & 81 & 3917 & 208 & 321 & 1055412 & 3636 & 179 & 303 & 981057\\
		\hline		
		\multicolumn{2}{|l|}{AlexNet (V:17)} & 1573  & 5     & 5     & 1714  & 91    & 141   & 3925  & 1650  & 83    & 139   & 37828 \\
		\multicolumn{2}{|l|}{VGG  \ \ \ \ \  (V:47)} & 1639  & 5     & 5     & 1787  & 95    & 147   & 113107 & 1719  & 86    & 144   & 108956 \\
		\hline
		\multicolumn{2}{|c|}{\multirow{2}{*}{Cmp.}} & 2984  & 59  & 60  & 3298  & 175   & 270   & 347524  & 3093  & 154   & 258   & 326188  \\
		     &  &       &       &       & 1     & 1     & 1     & 1     & 93.4\% & 86.8\% & 95.3\% & 94.0\% \\
		\hline
	\end{tabular}%
	\label{tab:init_moduleshape_cmp}%
\end{table}%

\subsubsection{\textbf{Comparison of generating the solution of task modules partition, schedule and floorplan }}
\label{sec:init_shape_cmp}	
	The method $Int\_PSF$ in \cite{chen2018integrated}  can be used to explore the solution space to search the optimal solution of task modules partitioning, scheduling, and floorplanning on the PDRS with modeling the chip with homogeneous resource, CLB.  
	Our  complete workflow explores more possibility of floorplan solutions,  thus making the solution with greater probability tends to optimize the scheduling time and communication cost. 
	The complete workflow  makes the design use more resources on the chip with trying to let the whole design of PDRS satisfy the boundary constraints, thus reducing the scheduling time of the entire design and decreasing the communication cost. Therefore, we compare the solutions generated by these two exploration methods.  In this comparative experiment, we run the two exploration methods  10 times independently to make a comparison and show the experiment results in Table \ref{tab:cmp_solution}.

\begin{table}[htbp]
	\centering\small
	\caption{Comparison of generating the solution of task modules partition, schedule and floorplan}
	\resizebox{\textwidth}{!}{
	\begin{tabular}{|l|llllllll|llllllll|}
		\hline
		 \multicolumn{1}{|c|}{\multirow{2}{*}{Imp.}} & \multicolumn{1}{c}{} & \multicolumn{7}{c|}{Int\_PSF\cite{chen2018integrated}} &  \multicolumn{1}{c}{} & \multicolumn{6}{c}{Our complete workflow } &  \\ \cline{3-17} 
		 \cline{2-17}
		 \multicolumn{1}{|c|}{} & $SchB$ & $SchA$ & $CC$ & \multicolumn{3}{c}{$RRT$} & $succ$ & $RT$ & $SchA$ & $SchM$ & $CC$ & \multicolumn{3}{c}{$RRT$} & $succ$ & $RT$ \\ 
			\hline
			t10-1 & 160.1 & 173.9 & 25147 & \multicolumn{3}{c}{{[}28.9\%, 27.1\%,30.2\%{]}} & 100\% & 2.4 & 155.1 & 155.1 & 21057 & \multicolumn{3}{c}{{[}55.7\%, 53.7\%,58.6\%{]}} & 100\% & 2.53 \\
			t10-2 & 265.4 & 267 & 36153 & \multicolumn{3}{c}{{[}28.6\%, 28.2\%,29.5\%{]}} & 100\% & 2.3 & 265.4 & 265.4 & 23646 & \multicolumn{3}{c}{{[}62.0\%, 61.7\%,63.1\%{]}} & 100\% & 1.98 \\
			t10-3 & 365.1 & 368.4 & 60296 & \multicolumn{3}{c}{{[}24.7\%, 24.3\%,23.1\%{]}} & 100\% & 1.9 & 365.1 & 365.1 & 54662 & \multicolumn{3}{c}{{[}67.1\%, 66.9\%,67.3\%{]}} & 100\% & 2.04 \\
			\hline
			t30-1 & 731.7 & 732.5 & 464762 & \multicolumn{3}{c}{{[}27.5\%, 26.4\%,29.0\%{]}} & 100\% & 12.3 & 731.7 & 731.7 & 473692 & \multicolumn{3}{c}{{[}53.0\%, 51.8\%,57.7\%{]}} & 100\% & 11.27 \\
			t30-2 & 1456.3 & 1456.3 & 5471700 & \multicolumn{3}{c}{{[}42.5\%, 42.0\%,45.9\%{]}} & 100\% & 9.7 & 1456.3 & 1456.3 & 5378078 & \multicolumn{3}{c}{{[}70.5\%, 70.5\%,72.0\%{]}} & 100\% & 8.97 \\
			t30-3 & 797 & 810.3 & 530415 & \multicolumn{3}{c}{{[}32.8\%, 31.2\%,30.3\%{]}} & 100\% & 9.4 & 792.2 & 792.2 & 515464 & \multicolumn{3}{c}{{[}54.7\%, 54.8\%,55.6\%{]}} & 90\% & 8.92 \\
			\hline
			t50-1 & 601.1 & 601.1 & 421508 & \multicolumn{3}{c}{{[}43.4\%, 42.7\%,46.5\%{]}} & 100\% & 25.6 & 601.1 & 601.1 & 376342 & \multicolumn{3}{c}{{[}54.1\%, 54.3\%,54.5\%{]}} & 100\% & 20.55 \\
			t50-2 & 326.9 & 333.8 & 187369 & \multicolumn{3}{c}{{[}73.5\%, 73.8\%,79.2\%{]}} & 100\% & 14.9 & 326.3 & 330.3 & 172231 & \multicolumn{3}{c}{{[}74.9\%, 75.1\%,75.1\%{]}} & 100\% & 20.4 \\
			t50-3 & 1080.8 & 1093.7 & 3973633 & \multicolumn{3}{c}{{[}44.9\%, 43.9\%,42.0\%{]}} & 100\% & 19.7 & 598.6 & 684.2 & 2949261 & \multicolumn{3}{c}{{[}83.1\%, 83.3\%,83.3\%{]}} & 70\% & 19.43 \\
			\hline
			t100-1 & 993.5 & 1069.4 & 6532062 & \multicolumn{3}{c}{{[}68.0\%, 67.7\%,73.5\%{]}} & 100\% & 70.9 & 897.5 & 975.7 & 5278860 & \multicolumn{3}{c}{{[}82.3\%, 82.8\%,82.8\%{]}} & 100\% & 77.66 \\
			t100-2 & 1440 & 1512.4 & 7665218 & \multicolumn{3}{c}{{[}60.2\%, 60.1\%,64.6\%{]}} & 100\% & 69 & 1282.3 & 1341.9 & 7160849 & \multicolumn{3}{c}{{[}72.9\%, 73.7\%,73.7\%{]}} & 100\% & 77.21 \\
			t100-3 & 2309.2 & 2317.2 & 29033640 & \multicolumn{3}{c}{{[}47.0\%, 46.2\%,43.7\%{]}} & 100\% & 69 & 1192.4 & 1302.3 & 19075710 & \multicolumn{3}{c}{{[}90.4\%, 90.5\%,90.6\%{]}} & 100\% & 72.7 \\
			\hline
			t200-1 & 3450 & 3404.6 & 1.96E+08 & \multicolumn{3}{c}{{[}82.2\%, 79.5\%,88.0\%{]}} & 100\% & 219.6 & 3051.5 & 3759.2 & 1.87E+08 & \multicolumn{3}{c}{{[}82.3\%, 82.3\%,82.4\%{]}} & 100\% & 564.7 \\
			t200-2 & 5377.1 & 5245.6 & 99734250 & \multicolumn{3}{c}{{[}61.5\%, 61.1\%,65.5\%{]}} & 100\% & 267.5 & 4658.1 & 5067.4 & 93725770 & \multicolumn{3}{c}{{[}73.6\%, 73.4\%,74.7\%{]}} & 100\% & 610.4 \\
			t200-3 & 2809.7 & 2788.2 & 10030710 & \multicolumn{3}{c}{{[}46.6\%, 46.1\%,43.2\%{]}} & 100\% & 258.2 & 1546.3 & 1743.1 & 7174263 & \multicolumn{3}{c}{{[}88.1\%, 88.3\%,88.6\%{]}} & 100\% & 523.9 \\
			\hline
			AlexNet & 50.1 & 50.3 & 72789 & \multicolumn{3}{c}{{[}33.3\%, 33.0\%,33.8\%{]}} & 100\% & 5.8 & 50.1 & 50.1 & 70066 & \multicolumn{3}{c}{{[}59.0\%, 58.0\%,62.8\%{]}} & 100\% & 5.2 \\
			\hline
			VGG & 503.5 & 505.1 & 3066295 & \multicolumn{3}{c}{{[}35.2\%, 34.5\%,36.3\%{]}} & 100\% & 20.9 & 503.5 & 504.4 & 2737673 & \multicolumn{3}{c}{{[}69.4\%, 69.2\%,72.5\%{]}} & 100\% & 25.4 \\
			\hline
			Cmp. & 1336.3 &1337.0  &21361003  & \multicolumn{3}{c}{{[}45.9\%, 45.2\%,47.3\%{]}} & 100\% & - &1086.7  & 1183.9 & 19527196 & \multicolumn{3}{c}{{[}70.2\%, 70.0\%,71.5\%{]}} & 98\% & -\\
				& 1     & 1     & 1     & \multicolumn{3}{c}{-}  &  -     &    -   & 81.3\% & 88.5\% & 91.4\% & \multicolumn{3}{c}{+24.3\%, +24.8\%, +24.2\%}  &     -  & - \\
			\hline
	\end{tabular}%
	}
	\label{tab:cmp_solution}%
\end{table}

	In Table \ref{tab:cmp_solution}, the column $SchB$ and $SchA$ is the best and the average value of scheduling time,  respectively. And their unit is $ms$.  The column $CC$ is the average  communication cost and is calculated by the method in \cite{chen2018integrated}. The column $RRT$ means the resources reuse rate and is calculated by Equation (\ref{eq:RRT}). In Equation (\ref{eq:RRT}), $Schedule_{time}$ is the schedule time of the PDRS, $Resource_i^{chip}$ means the number of the corresponding resource that the chip has, the $Resource_i^{drr_j}$ means the number of the corresponding resource that the reconfigurable region $drr_j$ has, the columns $conf_j$ and $exec_j$ mean the total configuration time and total execution time of the reconfigurable region $drr_j$, respectively, and the column $RT$ means the run time of the exploration process.
	These value are calculated from the experiment results that has feasible floorplan in the 10 independent experiment results. And the column $succ$ means the success rate of floorplan.

	\begin{equation}
		\label{eq:RRT}
		RRT_{i,i=CLB,BRAM,DSP} = \frac{\sum_{j}^{N_{DRR}}{Resource_i^{drr_j}}\times(conf_{drr_j} + exec_{drr_j})}{Schedule_{time}*Resource_i^{chip}}
	\end{equation}
	
	As shown in Table \ref{tab:cmp_solution}, in these experiments that have feasible floorplan, our complete workflow  can fully utilize the on-chip resources, improve the resource reuse rate, and thus find a smaller scheduling time and communication cost.
	Compared to the method $Int\_PSF$ in \cite{chen2018integrated}, the best schedule time and the average schedule time can be 18.7\% and 11.5\% better, respectively, the communication cost can be 8.6\% better, the resources reuse rate of CLB, BRAM, and DSP is 24.3\%, 24.8\%, and 24.2\% better, respectively.  

\subsubsection{\textbf{Comparison of reselecting task modules shape }}
\label{sec:diff_method_cmp}

	The floorplan of the solution generated by the exploration process may be infeasible due to the boundary constraint.   
	We use the ILP model to constrain the position relationship among task modules that the partition, schedule, and floorplan have been determined, and reselect the shape of task modules so that improve the success rate of the floorplan.  It is worth noting that, because some infeasible solutions can be feasible by post-optimization, it is possible to find a smaller schedule time just like the benchmark $t200-2$.
	 The experimental results are shown as Table \ref{tab:succ_cmp}, by solving the ILP model, the floorplan success rate can be improved to 98\%, compared with the solutions generated by the exploration process.  

\begin{table}[htbp]
	\centering\small
	\caption{Comparison of reselecting task modules shape}
	\begin{tabular}{cl|ll|lll}
		\hline
		\multirow{2}{*}{Bench.} & \multicolumn{1}{c}{\multirow{2}{*}{Imp.}} & \multicolumn{2}{c}{Solution}  & \multicolumn{2}{|c}{After ILP} \\
		\cline{3-7}
		& \multicolumn{1}{c}{} &$SchB$& succ   & $SchB$  & succ   & RT    \\
		\hline
		\multirow{3}{*}{t10}
		 & 1 & 155.1 & 100\% & 155.1 & 100\% & 2.5 \\
		& 2 & 265.4 & 100\% & 265.4 & 100\% & 1.9 \\
		& 3 & 365.1 & 100\% & 365.1 & 100\% & 2 \\
		\hline
		\multirow{3}{*}{t30} 
		& 1 & 731.7 & 100\% & 731.7 & 100\% & 11.1 \\
		& 2 & 1456.3 & 50\% & 1456.3 & 100\% & 8.6 \\
		& 3 & 792.2 & 90\% & 792.2 & 90\% & 8.8 \\
		\hline
		\multirow{3}{*}{t50}
		 & 1 & 601.1 & 80\% & 601.1 & 100\% & 19.4 \\
		& 2 & 326.3 & 60\% & 326.3 & 100\% & 16.7 \\
		& 3 & 598.6 & 70\% & 598.6 & 70\% & 18.7 \\
		\hline
		\multirow{3}{*}{t100} 
		& 1 & 897.5 & 70\% & 897.5 & 100\% & 64.5 \\
		& 2 & 1282.3 & 100\% & 1282.3 & 100\% & 64.3 \\
		& 3 & 1192.4 & 100\% & 1192.4 & 100\% & 63.4 \\
		\hline
		\multirow{3}{*}{t200} 
		& 1 & 3051.5 & 30\% & 3051.5 & 100\% & 160.4 \\
		& 2 & 4672.8 & 80\% & *4658.1 & 100\% & 202.4 \\
		& 3 & 1546.3 & 100\% & 1546.3 & 100\% & 237.6 \\
		\hline
		\multicolumn{2}{c}{AlexNet} & 50.1 & 100\% & 50.1 & 100\% & \textless{}0.1 \\
		\hline
		\multicolumn{2}{c}{VGG} & 503.5 & 100\% & 503.5 & 100\% & 1	\\
		\hline
		\multicolumn{2}{c}{Cmp.} &    -   & \textbf{84\%} & - & \textbf{98\%} &-  \\
		\hline
	\end{tabular}%
	\label{tab:succ_cmp}
\end{table}

\section{Conclusions}\label{sec:conclusion}

	In this paper, we propose the complete workflow to solve the problem of task modules partitioning, scheduling, and floorplanning on heterogeneous resources based on FPGA-PDRS.
	This complete workflow includes three parts, pre-processing,  the exploration of task modules partitioning, scheduling, floorplanning, and post-optimization.  
	Firstly,  in this pre-processing process, we propose the strategy of generating the list of task  modules  candidate shapes according to resources requirements of each task module. 
	Then we change the algorithm in our previous work by  changing the process of exploring solution space and adding the cost of heterogeneous resource utilization to generate the solution of task modules partitioning, scheduling and floorplanning  and integrate this algorithm into the complete workflow.
	Finally, based on the solution generated by the exploration process, we build the ILP model to reselect task modules shape from the  candidate shapes list to improve the success rate of floorplan.
	Experimental results show that, compared with state-of-the-art work, the  proposed complete workflow can improve performance by 18.7\%, reduce communication cost by 8.6\%. And based on the solution generated by the exploration process,  the post-optimization can optimize the floorplan success rate to 98\% with a 14\% improvement.

\section{ACKNOWLEDGMENTS}
	This work was supported in part by the National Key R\&D Program of China under grant No. 2019YFB2204800, in part by National Natural Science Foundation of China (NSFC) under grant Nos. 61874102, U19A2074, 61931008, and 62141415, in part by CAS Project for Young Scientists in Basic Research under grant No. YSBR-029k, in part by the Strategic Priority Research Program of Chinese Academy of Sciences, Grant No. XDB44000000. The authors would like to thank Information Science Laboratory Center of USTC for the hardware \& software services.

\balance
{
    \bibliographystyle{IEEEtran}
    \bibliography{./ref/Top-sim,./ref/reference,./ref/Software,./ref/ALG}
}

\end{document}